\documentclass{jps-cp}
\usepackage{txfonts} 

\title{Structural Response to the Hidden Charge Order in Sr$_\text{3}$Fe$_\text{2}$O$_\text{7}$}

\author{Darren C.\ \textsc{Peets}$^{1}$, Nikita D.\ \textsc{Andriushin}$^{1}$, Tobias \textsc{Ritschel}$^{1}$, Morgan C.\ \textsc{Allison}$^{1}$, Nikolai S.\ \textsc{Pavlovskii}$^{1}$, Anton A.\ Kulbakov$^{1}$, Alexandre \textsc{Ivanov}$^{2}$, Jose R.\ \textsc{Linares Mardegan}$^{3}$, Pablo J.\ \textsc{Bereciartua}$^{3}$, Sonia \textsc{Francoual}$^{3}$, Jochen \textsc{Geck}$^{1}$, and Dmytro S.\ \textsc{Inosov}$^{1,4}$}

\inst{$^{1}$Institut f\"ur Festk\"orper und Materialphysik, Technische Universit\"at Dresden, H\"ackelstr.\ 3, Dresden 01062, Germany\\
  $^{2}$Institut Laue-Langevin, 71 Avenue des Martyrs, CS 20156, CEDEX 9, 38042 Grenoble, France\\
  $^3$Deutsches Elektronen-Synchrotron DESY, Hamburg 22603, Germany\\
$^4$W\"urzburg-Dresden Cluster of Excellence on Complexity and Topology in Quantum Matter\,---\,{\slshape ct.qmat}, Technische Universit\"at Dresden, Dresden 01062, Germany}

\email{darren.peets@tu-dresden.de}

\recdate{September 5, 2026}

\abst{We previously reported ``hidden'' charge order below $\sim$330\,K in the centrosymmetric skyrmion host Sr$_3$Fe$_2$O$_7$ [J.-H.\ Kim {\itshape et al}., Phys.\ Rev.\ Lett.\ {\bfseries 127}, 097203 (2021)], finally identifying the charge disproportionation and localization transition that had been first seen in the 1960s.  We detected this through resonant x-ray scattering, which directly probes electron density, but is only indirectly sensitive to the lattice structure.  This left open the question of whether the charge-ordering transition has a true structural component or consists merely of the reorganization of electrons.  Neutrons probe the lattice directly, and we now show, based on neutron data, that the charge order is accompanied by a clear structural modification.}

\begin{document}
\maketitle

\section{Introduction}

Sr$_3$Fe$_2$O$_7$ is a multiple-$Q$ helical material below $T_\text{N}=112$\,K, hosting a novel helix+sine ground state, a spin meta-cholesteric state at intermediate temperatures, and an equal-$Q$ sine+sine phase at high temperatures, as well as a variety of other phases in applied fields\cite{Andriushin2023,Andriushin2024}.  The complex magnetic order and rich magnetic phase diagram are strikingly similar to those found in the cubic perovskite analogue SrFeO$_3$\cite{Ishiwata2020,Andriushin2025}.  Of note, SrFeO$_3$ is a metal, while Sr$_3$Fe$_2$O$_7$ is an insulator.  Indeed, the proximity to a metal-insulator transition may be responsible for the helical magnetism\cite{Kim2014}.  However, for a long time there were questions around the crystal structure underlying this magnetism.

Sr$_3$Fe$_2$O$_7$ undergoes a charge-disproportionation and localization transition around 330\,K.  The first indications of the charge disproportionation came from $^{57}$Fe Moessbauer spectroscopy as early as the 1960s\cite{Gallagher1966}, where two distinct iron valences with roughly equal populations were visible.  However, the crystal structure of Sr$_3$Fe$_2$O$_7$ showed no deviations from $I4/mmm$, which only has one Fe site.  Using elastic x-ray scattering near the Fe K absorption edge, we showed that the material exhibits well-ordered two-dimensional checkerboard charge order described by the space group $Bmmb$ (an alternate setting of $Cmcm$, No.\ 63). However, the three-dimensional stacking of this charge order is frustrated by the body centering in $I4/mmm$ (in $Bmmb$, the $b$ glide plane plays a similar role in relating the bilayer in the middle of the unit cell to that at the top/bottom)\cite{Kim2021}.  This doubles the unit cell in the $ab$ plane, leading to new structural Bragg peaks, and rotates it by 45$^\circ$\,---\,here we use the $I4/mmm$ setting unless otherwise stated.  The out-of-plane correlations extend over only approximately one unit cell (two bilayers), broadening the new Bragg peaks significantly along $L$. This in turn reduces their peak intensity below the detection limit for the neutron and non-resonant x-ray diffraction measurements we and other groups had previously performed.

This structurally frustrated charge order is particularly interesting because it is effectively a hidden order, with strong signatures in the specific heat and resistivity at the transition, but with a nearly undetectable structural component.  Many of the other hidden-order materials also crystallize in $I4/mmm$ or similar body-centred space groups\cite{Mydosh2011,Mydosh2020,Cao2016}, and it is possible that diffraction signatures of the hidden order may have been missed due to similar body-centring-derived frustration.  

Since our unmasking of the structural hidden order in Sr$_3$Fe$_2$O$_7$ was based on x-ray techniques, which probe charge distribution, it remained unclear whether the atomic nuclei also shift at the transition.  We show here, based on neutron scattering, which is sensitive to nuclear positions, that they do.

\section{Experimental}

Single crystals of Sr$_3$Fe$_2$O$_7$ were grown by the floating-zone technique as described elsewhere\cite{Maljuk2004,Peets2013}.  The resulting crystals of Sr$_3$Fe$_2$O$_{6-\delta}$ were then fully oxygenated to Sr$_3$Fe$_2$O$_{7}$ under 5\,kbar of oxygen at 450$^\circ$C, with slow cooling over the course of a week to maximize the oxygen content\cite{Kim2021}.

Resonant elastic x-ray scattering (REXS)\cite{Fink2013,Lorenzo2012} was performed on a single crystal of Sr$_3$Fe$_2$O$_7$ on the low-energy side of the Fe K edge (photon energy 7112\,eV) at beamline P09 at PETRA-III, DESY, in Hamburg, Germany\cite{Strempfer2013}. The incoming polarization was perpendicular to the scattering plane, and for most of the data the outgoing polarization was not analyzed.  Where $\sigma$--$\sigma$ and $\sigma$--$\pi$ polarizations were compared to determine whether the peaks contained magnetic intensity, a Cu [110] analyzer crystal was used.

\begin{figure}[b!]
  \includegraphics[width=\textwidth]{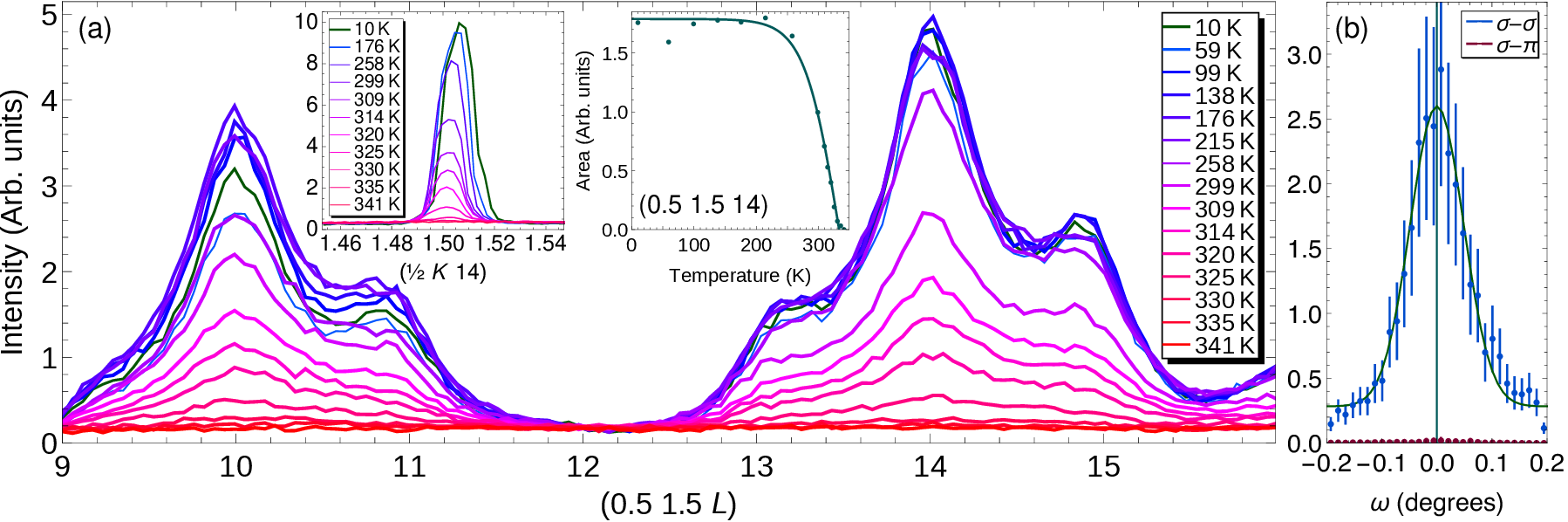}
  \caption{\label{REXS2}(a) A cut along $L$ through structural peaks arising from the charge disproportionation at $\left(\frac12\,\frac32\,L\right)$ in our REXS data at various temperatures.  Left inset:  An in-plane cut through this along $K$ at $L=14$.  The right inset shows the temperature dependence of the integrated intensity at $\left(\frac12\,\frac32\,14\right)$, fit to a tanh curve. (b) Rocking curves of the $\left(\frac12\,\frac12\,13\right)$ peak for $\sigma$--$\sigma$ and $\sigma$--$\pi$ polarizations at 5.0\,K, with the former fit to a Gaussian.}
\end{figure}

Neutron scattering experiments were performed on the ThALES cold-neutron triple-axis spectrometer at the Institut Laue-Langevin (ILL) in Grenoble, France\cite{Boehm2008}, with the crystal mounted in the $[HHL]$ scattering plane based on the $I4/mmm$ setting.
The final neutron wavevector $k_f$ was set to 1.30\,\AA$^{-1}$, corresponding to 3.502\,meV, and elastic-scattering measurements were performed by setting the analyzer to the same energy.  Higher-order wavelength contamination was suppressed using a velocity selector.  The energy resolution was approximately 0.10\,meV FWHM.  Triple-axis spectrometers are optimized for measuring inelastic spectra which are several orders of magnitude weaker than ordinary Bragg reflections, affording us very high intensity to investigate weak peaks, and their energy analysis allows significantly reduced background.

\section{Resonant Elastic X-ray Scattering}

\begin{figure}[b!]
  \includegraphics[width=\textwidth]{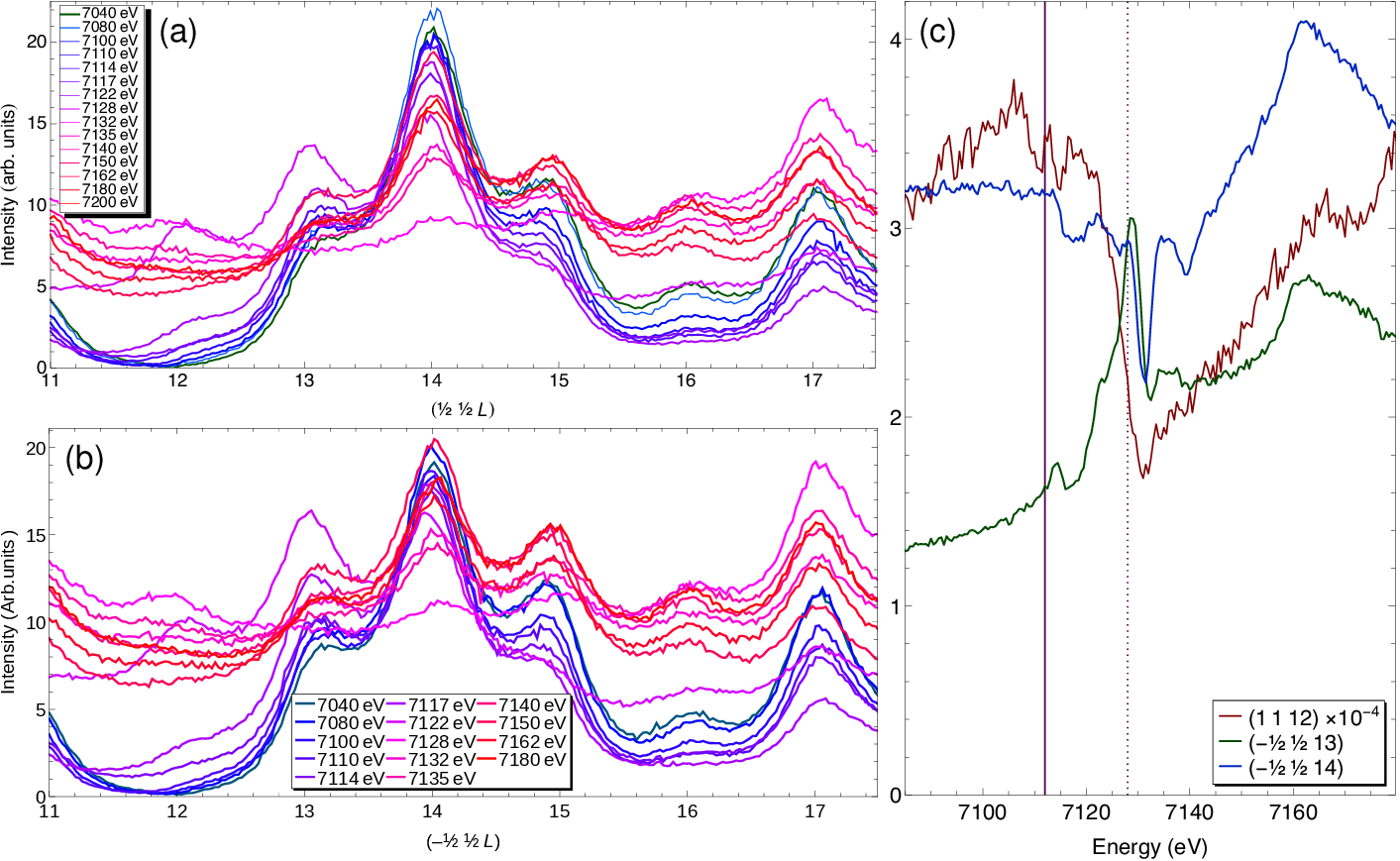}
  \caption{\label{Edep}Dependence of the resonant elastic x-ray spectra on photon energy.  (a) Scan along $L$ at $\left(\frac12\,\frac12\,L\right)$.  (b) Scan along $L$ at $\left(-\frac12\,\frac12\,L\right)$.  (c) Energy scans on three peaks, two of which are associated with the charge order.  The energies used in the previous study\,\cite{Kim2021} are marked with vertical lines; the data in Fig.~\ref{REXS2} were measured at 7112\,eV, marked by the solid line.}
\end{figure}

As we previously reported\cite{Kim2021}, REXS at the Fe K edge finds intensity at $\left(\frac12\,\frac12\,L\right)$ and equivalent positions in $I4/mmm$, the existence and extinctions of which demonstrate that the space group is actually $Bmmb$ below $\sim$330\,K.  Additional data are shown in Fig.~\ref{REXS2}(a), for cuts along $K$ (left inset) and $L$.  In plane (along $K$) the width is resolution limited, but along $L$ it is a large fraction of a reciprocal lattice unit.  The temperature dependence of the intensity of this peak is shown in the right inset\,---\,an empirical fit to a tanh curve gives a transition temperature of 333\,K.  A similar fit of the $\left(\frac12\,\frac12\,14\right)$ data in Ref.~\cite{Kim2021} gave a transition temperature of 329\,K.  It is worth noting that the intensity at this position is not obviously sensitive to the magnetic transition at $\sim$112\,K, while a comparison of $\sigma$--$\sigma$ and $\sigma$--$\pi$ polarizations [Fig.~\ref{REXS2}(b)] deep in the magnetically ordered state at 5.0\,K confirms that these peaks are electronic or structural in origin and not magnetic.

Figures~\ref{Edep}(a) and \ref{Edep}(b) show scans along $L$ at several different energies, at $\left(\frac12\,\frac12\,L\right)$ and $\left(-\frac12\,\frac12\,L\right)$, respectively.  One might expect that the intensities in Fig.~\ref{REXS2}(a) and in a similar scan at $\left(\frac12\,\frac12\,L\right)$ in our previous work\,\cite{Kim2021} are contained in an envelope which could be described by a standard bilayer structure factor.  However, these scans at different energies demonstrate that, as expected near absorption edges, the total structure factor is strongly and nontrivially energy dependent.

Figure~\ref{Edep}(c) shows energy scans through three distinct structural peaks:  $(1\,1\,12)$, which is allowed in $I4/mmm$; $\left(-\frac12\,\frac12\,13\right)$, a $Bmmb$ peak with $L$ odd; and $\left(-\frac12\,\frac12\,14\right)$, a $Bmmb$ peak with $L$ even.  The energy dependence of $(1\,1\,12)$ shows the Fe K edge inverted, as expected\,---\,when the material becomes more strongly absorbing at the resonance, fewer photons are diffracted.  The other two peaks, however, exhibit a series of peaks and valleys across the Fe K edge and are also very different from each other.  The modeling required to fully understand these features has not been performed, but they reflect the difference in valence, orbital occupation, and possibly symmetry between the two iron sites. 

Our REXS data demonstrate that charge order is present.  However, this transition is associated with charge redistribution, so a shift in the electron density does not automatically imply that the atomic positions change.  The charge redistribution could to some extent be independent of the positions of the atomic nuclei.  We thus sought to confirm this result with an independent measurement of the nuclear positions.

\section{Neutron-Scattering Results}

\begin{figure}[b!]
  \includegraphics[width=\textwidth]{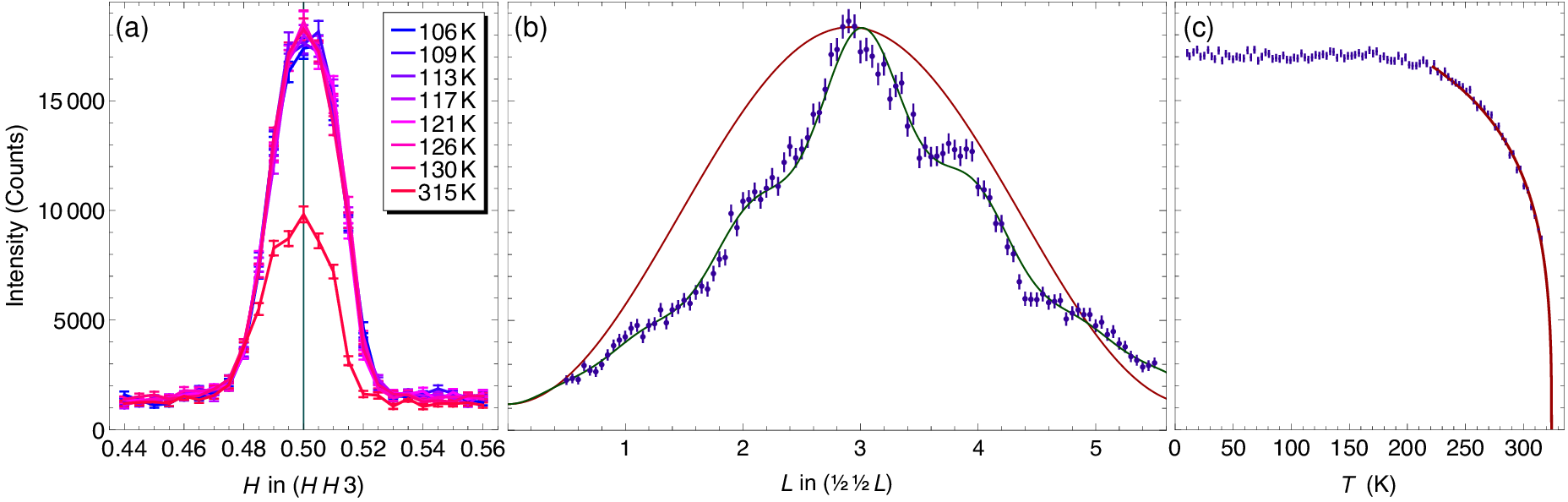}
  \caption{\label{Ldep}(a) Elastic scans through the $\left(\frac12\,\frac12\,3\right)$ peak in the $I4/mmm$ setting at several temperatures, from our inelastic ThALES neutron dataset.  (b) $L$-dependence at the $\left(\frac12\,\frac12\,L\right)$ position in the $I4/mmm$ setting in an elastic cut from our inelastic data.  As in the previously published x-ray data, the structural peaks arising from the charge order are broadened but present.  The solid red line represents schematically the bilayer structure factor of an isolated bilayer calculated based on the atomic positions in the $Bmmb$ structure, with a vertical offset corresponding to the background level in (a), while the solid green line is a fit to Gaussians at integer $L$ multiplied by a bilayer structure factor, with the same background added. (c) Temperature dependence of the neutron peak intensity at $\left(\frac12\,\frac12\,3\right)$.  The data on the approach to the transition were fit to an empirical tanh form as described in the text.}
\end{figure}

A cut along $L$ through $\left(\frac12\,\frac12\,L\right)$ is shown in Fig.~\ref{Ldep}(b), and a cut through $\left(\frac12\,\frac12\,3\right)$, the most intense peak, along $(H\,H\,3)$ is presented in Fig.~\ref{Ldep}(a).  As in the resonant elastic x-ray data, we see broad peaks with widths of a large fraction of a reciprocal lattice unit, modulated by the bilayer structure factor, which is plotted as a solid line in Fig.~\ref{Ldep}(b) for reference.  These peaks, when fit to Gaussians, have a full width at half maximum (FWHM) of 0.90(3) reciprocal lattice units, close to the width seen with x rays [0.98(3) for the data in Fig.~\ref{REXS2}(a)].  The resulting $c$-axis correlation lengths $\xi_c=\frac{1}{\text{FWHM}}$ are 1.11(3) and 1.02(3) unit cells, respectively.  For the data at $\left(\frac12\,\frac12\,L\right)$ presented in Ref.~\cite{Kim2021}, the FWHM was 0.70(2), giving a correlation length of 1.44(4) unit cells.  With two bilayers per unit cell, this is 2--3 bilayers.  While these correlation lengths indicate that consecutive bilayers are not completely uncorrelated, these are very short correlation lengths.  This is why the average crystal structure does not show any orthorhombic distortion and nonlocal probes that average over many bilayers struggle to see the structural distortions.

The data in Fig.~\ref{Ldep}(b) are fit to a sum of Gaussians in a bilayer envelope:
\begin{equation}\label{fit}
  Intensity = I_0 + \sin^2\left(\pi L \frac{2d}{c}\right)\sum_{L_i=0}^6 \sqrt{\frac2\pi} \frac{A_i}{w} e^{-\frac{2(L-L_i)^2}{w^2}}
\end{equation}
where $I_0$ is taken from fits to the data in Fig.~\ref{Ldep}(a) and all peaks are assumed to have the same width $w$, and we also plot a crude bilayer structure factor for an isolated bilayer, based on the $Bmmb$ atomic positions: 
\begin{equation}\label{envelope}
  I_0 + I_\text{BL}\sin^2\left(\pi L \frac{\sum_i n_i\sigma_i d_i}{c\sum n_i\sigma_i}\right)
\end{equation}
Here, $\sigma_i$ is the neutron scattering cross-section for atom $i$, $d_i/c$ is the distance along $c$ from that atom to the centre of the nearest bilayer in fractional coordinates, and $n_i$ is the number of such ions at that $c$ position.  The intensity ratios of the peaks in Fig.~\ref{Ldep} do not follow what would be predicted based on the $Bmmb$ structure, but are also not fully captured by the crude bilayer structure factor based on an isolated bilayer.  In particular, the $L=2$ peak in $Bmmb$ should be several times stronger than any other peak in this $L$ range.  The effective bilayer spacing $d/c$ in the $\sin^2$ in Eq.~\ref{fit} is 0.0803(6) in our fit, while the effective $d/c$ calculated from the published structure in Eq.~\ref{envelope} is 0.0858, suggesting that the structure within a bilayer could likely still use some minor fine-tuning.  That this $L$ scan is better described by an approximate bilayer structure factor than by the Bragg intensities that should be present reinforces the conclusion that the structural order is strongly frustrated and has an extremely short correlation length along $c$.

The temperature dependence of the $\left(\frac12\,\frac12\,3\right)$ peak intensity is plotted in Fig.~\ref{Ldep}(c), and parameterized with an empirical fit to the data above 205\,K of the form
\begin{equation}
  I = A \tanh\left[\frac{\pi}{2}\left(\frac{T_\text{CO}}{T}-1\right)^C\right].
\end{equation}
This fit extrapolates to a charge-order transition at an estimated $T_\text{CO}$\,=\,324(1)\,K with power $C=0.34(1)$, but we note that reliability is limited by the required extrapolation, and $T_\text{CO}$ should be viewed as consistent with the x-ray version.  As for the x-ray results above, the intensity is not sensitive to the magnetic transitions at $\sim$70, 106, and 112\,K, again confirming that this reflection is purely structural.  Since the magnetic propagation vector is $(0.142~0.142~1)$, no magnetic intensity is expected at $\left(\frac12\,\frac12\,3\right)$.

\section{Discussion and Conclusion}

Taking advantage of the high signal-to-noise ratio and reduced background available when using an inelastic spectrometer for elastic scans, we demonstrate that the charge order in Sr$_3$Fe$_2$O$_7$ also involves the nuclei.  As in the REXS data [Fig.~\ref{REXS2}(a)], the neutron data (Fig.~\ref{Ldep}) show peaks which are resolution-limited within the plane, but a significant fraction of a reciprocal lattice unit wide along $L$.  They are insensitive to the magnetic transitions, and disappear around $\sim$330\,K, consistent with the charge-order transition.  This demonstrates that the charge reorganization is indeed associated with a structural reorganization\,---\,the nuclei also participate.  As was found for the charge density, the shifts in nuclear positions are well ordered in plane but the stacking is only correlated on the length scale of a single unit cell.  The $L$-dependence of the peak intensity is better described by the structure factor of an isolated bilayer than by the Bragg intensities expected from the $Bmmb$ crystal structure.  This stacking disorder is due to structural frustration arising from the body centering.  The difference between the two iron sites in terms of valence and lowered symmetry is evidenced by the highly nontrivial energy dependence in Fig.~\ref{Edep}. 

\section{Acknowledgements}

The authors are grateful for experimental assistance from the group of M.\ Jansen.  This project was funded by the German Research Foundation (Deutsche Forschungsgemeinschaft, DFG) through individual grants PE~3318/3, IN~209/7, and IN~209/9; through projects C03 and C06 of the Collaborative Research Center SFB~1143 (Project No.\ 247310070); through the W\"urzburg-Dresden Cluster of Excellence on Complexity and Topology in Quantum Matter\,---\,{\slshape ct.qmat} (EXC~2147, Project No.\ 390858490); and through the Collaborative Research Center TRR~80 (Project No.\ 107745057). DCP was supported by the Chinese Academy of Sciences through 2018PM0036. The authors acknowledge the Institut Laue-Langevin, Grenoble, France, for providing neutron beam time. We acknowledge DESY (Hamburg, Germany), a member of the Helmholtz Association HGF, for the provision of experimental facilities for resonant elastic x-ray diffraction. Parts of this research were carried out at PETRA III. Data were collected using beamline P09 operated by DESY Photon Science. Beam time was allocated for proposal No.\ I-20190500.  Data associated with this manuscript are available online\cite{4-01-1698,OPARA-1516}.

\end{document}